# Performance Comparison Between MIMO and SISO Systems Based on Indoor Field Measurements

Shailesh Chaudhari[1], Jingy Hu[2], Babak Daneshrad[3]
Dept. of Electrical Engineering,
University of California, Los Angeles, CA-90095
[1]schaudhari@ucla.edu, [2] jingyi@silvustechnologies.com,
[3]babak@ee.ucla.edu

Jesse Chen
Silvus Technologies, Inc.
Los Angeles, CA-90024
jesse@silvustechnologies.com

*Abstract*— In this paper, we quantify performance gain achieved if SISO system is replaced with 4x4 MIMO in WLAN setting compatible with IEEE 802.11n standard. We compare throughput and power savings in MIMO by taking field measurements at various indoor locations. Measurements are validated with simulations that include different IEEE TGn channel models. For comparison between MIMO and SISO, we select the Modulation and Coding Scheme (MCS) which yields the highest throughput subject to QoS constraints (PER <10%). We show that throughput gain of 2.5x-3x and power saving of 5-15 dB are achievable in 4x4 MIMO system.

## I. INTRODUCTION

With the advent of numerous internet capable devices, the demands for higher throughput and lower transmit power have greatly increased. Multiple antenna system such as Multiple Input Multiple Output (MIMO) are well suited to achieve throughput gain and power saving over single antenna systems. MIMO systems can outperform Single Input Single Output (SISO) systems by utilizing spatial diversity and multiplexing techniques. At low SNRs, spatial diversity in MIMO results in lower BER than SISO system. On the other hand, at higher SNR, spatial multiplexing can be exploited to improve throughput. Therefore it is expected that for same throughput requirements, MIMO system should have lower transmit power requirements and for the same transmit power constraints it should have higher throughput.

There are number of theoretical studies in literature regarding MIMO and SISO comparison. However, very few articles quantify advantages of MIMO over SISO in 802.11n with indoor field measurements and simulations. In [1], MIMO and SISO systems are compared in outdoor military environment at 430 and 1380 MHz. Throughput gain of MIMO over SISO is measured. However the power saving in MIMO is not studied. Also frequencies used are not applicable for civilian applications. In [2], throughput vs SNR curves are presented for MIMO and SISO. However, power saving s in MIMO are not studied. Also the IEEE Wi-Fi standard was not considered during the comparison. Reference [3] studies performance of MIMO system in outdoor-to-indoor environment and shows that in MIMO system, different eigen modes are optimum at different SNR values. However performance gain over SISO is not discussed. In another effort [4], performance of 2x2 MIMO-OFDM was studied using reconfigurable antennas in indoor environment. In [5], MIMO system performance is measured in terms of throughput for 802.11n system at 2.4GHz and 5GHz. However comparison with SISO is absent. There are number of research articles studying effect of different antennas on MIMO system performance [6-8].

None of the above articles quantify the performance gain of MIMO over SISO in terms of throughput gain and power saving, which is an important aspect in practical commercial system design. In this paper, we present the performance comparison for 802.11n system using 4x4 MIMO and SISO setup based on field measurements in various indoor environments e.g. office area, conference room, building corridors. At each location we compute throughput gain in MIMO system for same transmit power and power saving for same throughput. Rest of the paper is organized as follows: section II describes testbed hardware and simulation setup. In section III, test environment is described. In Section IV, we present and analyze simulation and measurement results. Section V concludes the paper.

## II. TESTBED SETUP

### A. Hardware and Software Implementation

The hardware platform for measurement consists of Silvus Technologies SC3500 radio [10]. SC3500 is configurable MIMO-ODFM platform which was used at 2.492 GHz with 20 MHz bandwidth. Transmit power can be a set from the computer interface. We further adjust the transmit power using attenuators attached at each antenna feed. This in turn adjusts received SNR. In this measurement, equal power is allocated to all transmit antennas in MIMO. Also same total power was transmitted in MIMO and SISO modes. Therefore each antenna in MIMO transmits 6dB less power as compared to SISO antenna. In the testbed, the transmitter transmits MIMO and SISO data in round robin fashion. The receiver switches between MIMO and SISO modes per second. At receiver PC interface, received SNR, throughput and corresponding MCS is displayed. The MCS having the highest throughput with QoS constraints PER<10% is also displayed. Further information regarding the hardware and processing can be found in [10]. Transmitter section of the setup is shown in Fig. 1

End-to-end simulation of 802.11n MIMO-OFDM system was developed in MATLAB. The simulation setup included following RF impairments: frequency offset, clock offset, quantization noise at Tx and Rx and transmit power backoff.

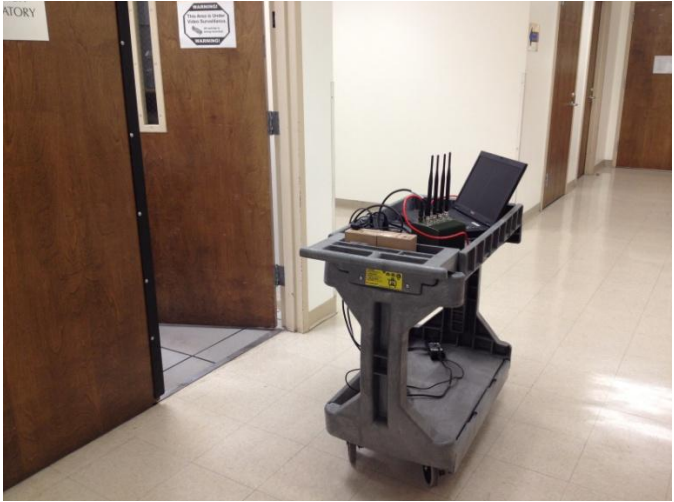

Fig. 1. Transmitter section of the setup

We simulate all MCS modes for MIMO and SISO and select the one with highest throughput with QoS constraints (PER<10%) for comparison between MIMO and SISO. The wireless channel in the simulation can be set to any one of TGn channel models [11].

*B. Calibration of testbed.*

In order to validate the results obtained after analyzing the measurements taken using above mentioned hardware and simulation engine, we calibrate the testbed in controlled environment where we can predict the measured results. For this we connect transmitter and receiver with wires to obtain AWGN channel. The received SNR is controlled by attenuators connected before the transmit antennas. Using this setup PER vs SNR curve is plotted and these results are compared with simulated results as shown in Fig. 2.

## III. TEST ENVIRONMENT

Measurements are taken at various locations at UCLA Engineering IV building. At each location, transmitter was kept in a fixed position and receiver was moved to different positions. During the measurement, both transmitter and receiver were stationary. Throughput vs SNR data points are collected for each location. Based on these data points, a third

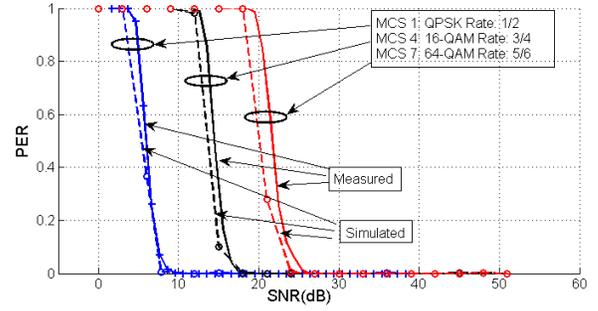

Fig. 2. Calibration of testbed.

order curve was plotted for further analysis.

*A. Location 1: Corridors*

The first set of measurements was carried out in corridors in Engineering IV building on the 5$^{th}$ floor. The corridor width is 2.4m. There number of doors leading to conference rooms in the side walls of the corridor.

*B. Location 2: Cubicle Area (53-138 Engr IV)*

This location has typical office cubicles with walkways between them. The area is surrounded by office rooms.

*C. Location 3: Farady Room (67-124 Engr IV)*

Faraday room is a rectangular conference room on 6$^{th}$ floor of Engineering IV. It is of size 9.2x3.5 m with two long desks and chairs around them.

*D. Location 4: Silvus Technologies Office*

This is a typical office location in a high rise building with number of conference and office rooms.

Layout of these locations along with positions of transmitter and receiver is shown in Appendix. The collected data points are shown in Fig. 3.

## IV. EXPERIMENTAL RESULTS

Instead of comparing MIMO and SISO performance separately at each location, all the data points collected at four locations are plotted on a single graph of Throughput vs SNR.

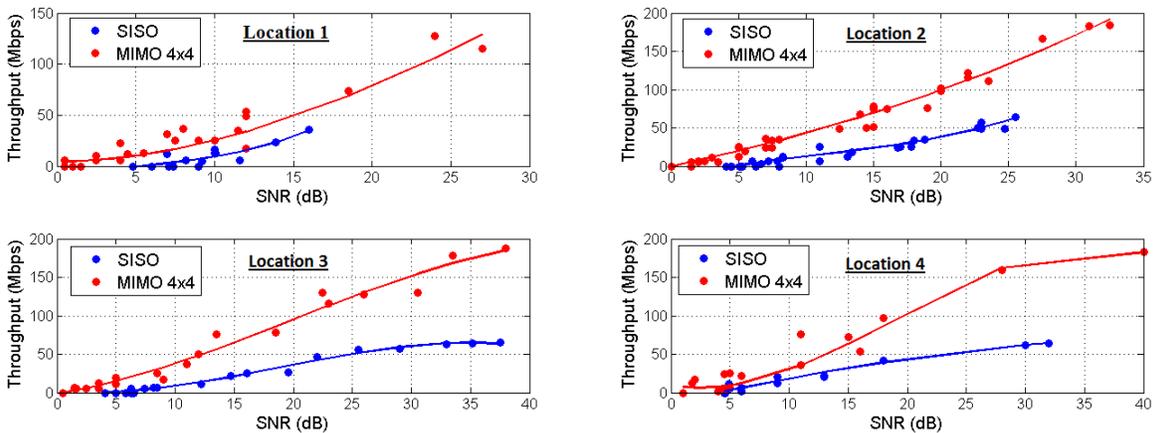

Fig. 3. Collected data points along with the best fit curve at Location 1, 2, 3 and 4 respectively

As seen from Fig. 4 all the data points can be approximated with a single third order curve for MIMO and SISO. We use these curves for further analysis. Data points collected at locations 2 and 3 lie in low SNR region, whereas points collected at location 1 lie in high SNR region.

*A. Comparison with Simulated Results*

We simulate 802.11n MIMO and SISO system with channel models A through E as described in [11]. The delay spread for simulated channels is 0, 15, 30, 50, 100 ns respectively. The third order curve is drawn for each simulation. According to 802.11n standards, maximum throughput is limited to 65 and 260 Mbps for SISO and MIMO respectively.

Further we compute the throughput gain and power saving in MIMO system. Throughput gain is defined as ratio of MIMO throughput to that of SISO at same transmit power. Since total transmit power was equal in MIMO and SISO, throughput gain can be calculated as ratio of throughputs in MIMO and SISO at equal received SNR. Similarly power saving in MIMO for fixed required throughput can be computed by taking difference between received SNR in MIMO and received SNR in SISO for that throughput. The simulation results are shown in Fig. 5.

It is observed that simulations with channel E (delay spread 100 ns) are better match with the measurements as compared to other channel models. It is important to note that the delay spread of wireless channel at these locations as measured in [12] was 50ns in cubicle area and approximately 30 ns in corridors.

From Fig. 5 we can see that throughput gain in MIMO is high (>4x) at low SNR values (<7 dB). This is due to the fact that SISO throughput is very small. After 30dB SNR, throughput gain curve changes shape since SISO throughput starts saturating at these SNR values. Measured power saving curve starts at 5 dB on vertical axis. This is attributed to approximately zero throughput of SISO for SNR < 5 dB. Power saving increases from 5 dB to 15 dB as required throughput is increased to 65 Mbps. This is the highest achievable throughput in 802.11n with SISO. Hence the curve is vertical at 65 Mbps.

*B. Comparison with Capacity*

In this section, we compare the throughput gain and power saving statistics calculated from channel capacity curve. Channel capacity for MIMO is calculated as follows:

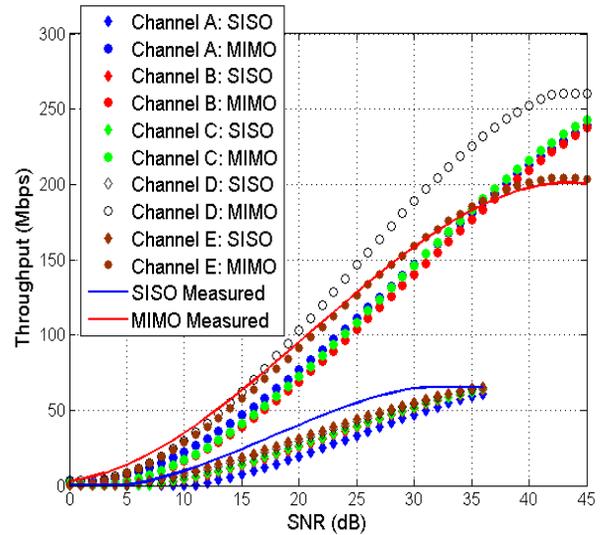

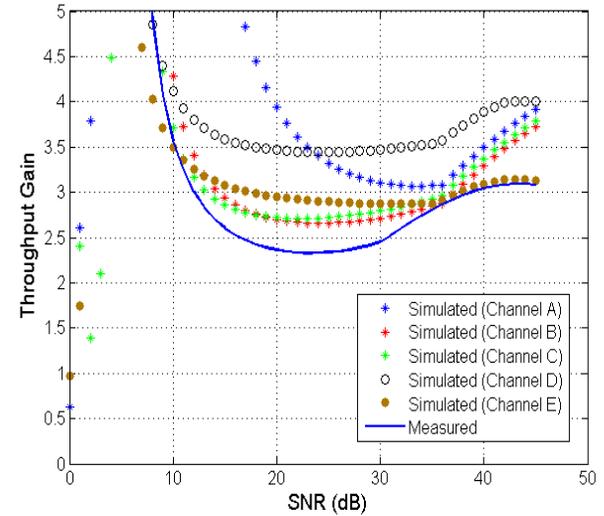

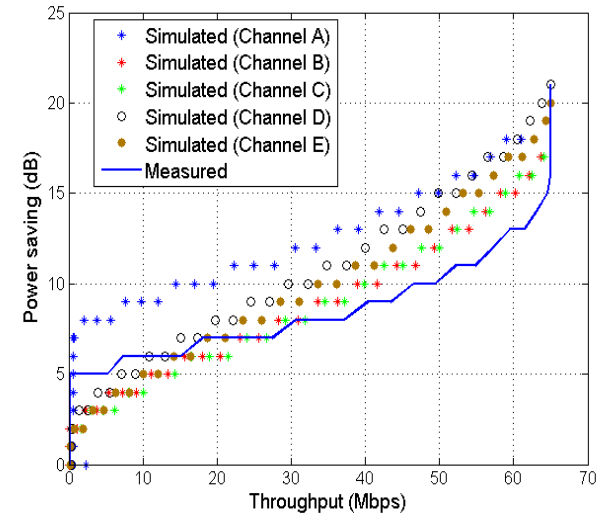

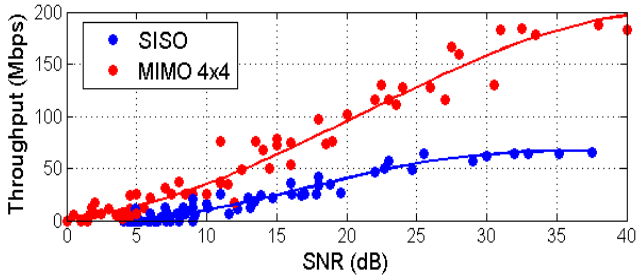

Fig. 4. Collected data points along with the best fit curve at Locations 1, 2, 3 and 4.

Fig. 5. Comparison of measured data with simulations: Throughput vs SNR, Throughput gain vs SNR and Power Saving vs SNR

$$C = B \sum_{i=1}^{4} log_2(1 + \frac{P_{tx}\lambda_i^2}{4 N_0}) \qquad (1)$$

Where B is system bandwidth, $P_{tx}$ is total transmit power, $\lambda_i$ is eigen mode and $N_0$ is noise variance. SISO capacity is obtained by replacing $\frac{P_{tx}\lambda_i^2}{4 N_0}$ with $\frac{P_{tx}}{N_0}$ and removing the summation over *i*.

As earlier, throughput gain and power saving is calculated this time using capacity curve. The results are shown in Fig. 6. We can see that Throughput vs SNR curve for capacity calculations is logarithmic whereas the measured and simulated curve is S-shaped. As expected the results obtained from capacity alone do not match measured and simulated results. Throughput gain calculated from capacity is constant at 3.2x and Power saving increases exponentially and is lower than measured values. Observations in this experiment are summarized in Table 1.

## V. CONCLUSION

In this paper, we quantified benefits of 4x4 MIMO system over SISO in terms of throughput gain and power savings. Experiments were carried out in indoor environment. Throughput gain of 2.5x-3x is expected if received SNR>12 dB. For lower SNR (<10dB), higher throughput gain is obtained. Power saving in MIMO system increased from 5-15 dB as required throughput reaches 65 Mbps in 802.11n system. Measurements were compared with simulated results with various TGn channel models. It is observed that the throughput vs SNR curves for MIMO and SISO closely match with simulations with channel model E with 100 ns delay spread. Therefore throughput gain and power savings can be predicted by simulating 802.11n system with this channel model.

We also compared our measured results with capacity curves. Throughput gain and power saving cannot be accurately predicted by capacity calculations alone. Throughput gains measured at low SNR values are much higher than those predicted by capacity. Similarly power saving values for all possible throughputs are higher than those obtained from capacity calculations. We have observed that the throughput vs SNR curve for measured and simulated results for MIMO and SISO is S-shaped. Further investigations are required in order to explain the behavior of curves obtained in this experiment.

TABLE I. PERFORMANCE GAINS IN 4x4 MIMO

|  | *Measured* | *Simulated* | *Computed from Capacity* |
|---|---|---|---|
| Throughput Gain for SNR > 12dB | 2.5x-3x | ~3x | 3.2x |
| Throughput Gain for SNR < 10dB | >4x | >4x | 3.2x |
| Power Saving | Upto 15 dB | Upto 20 dB | Upto 12 dB |

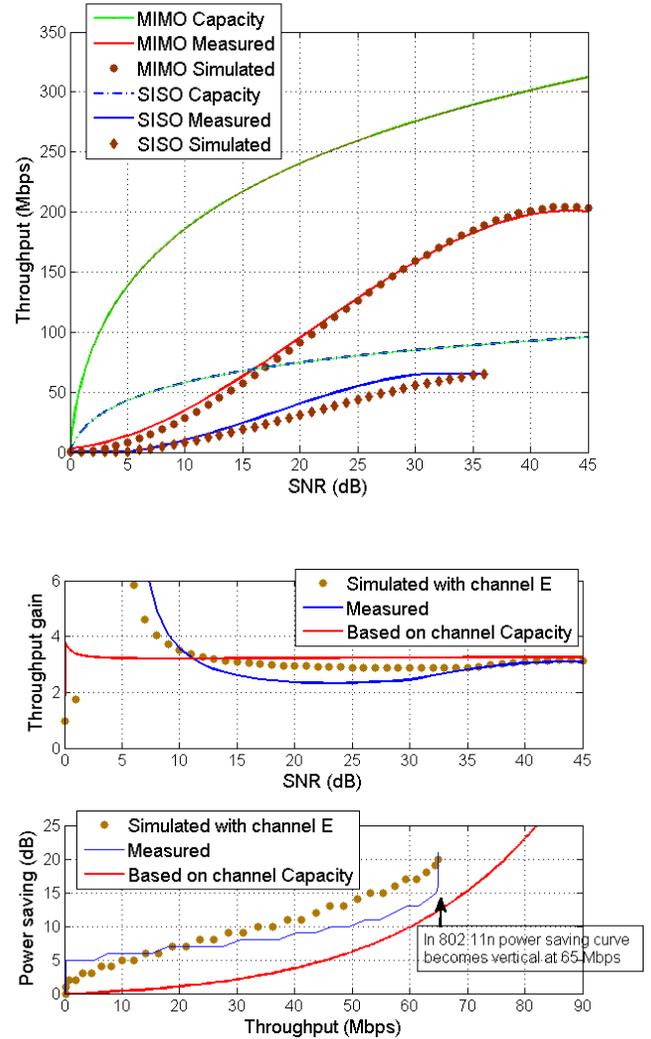

Fig. 6. Comparison with capacity

APPENDIX

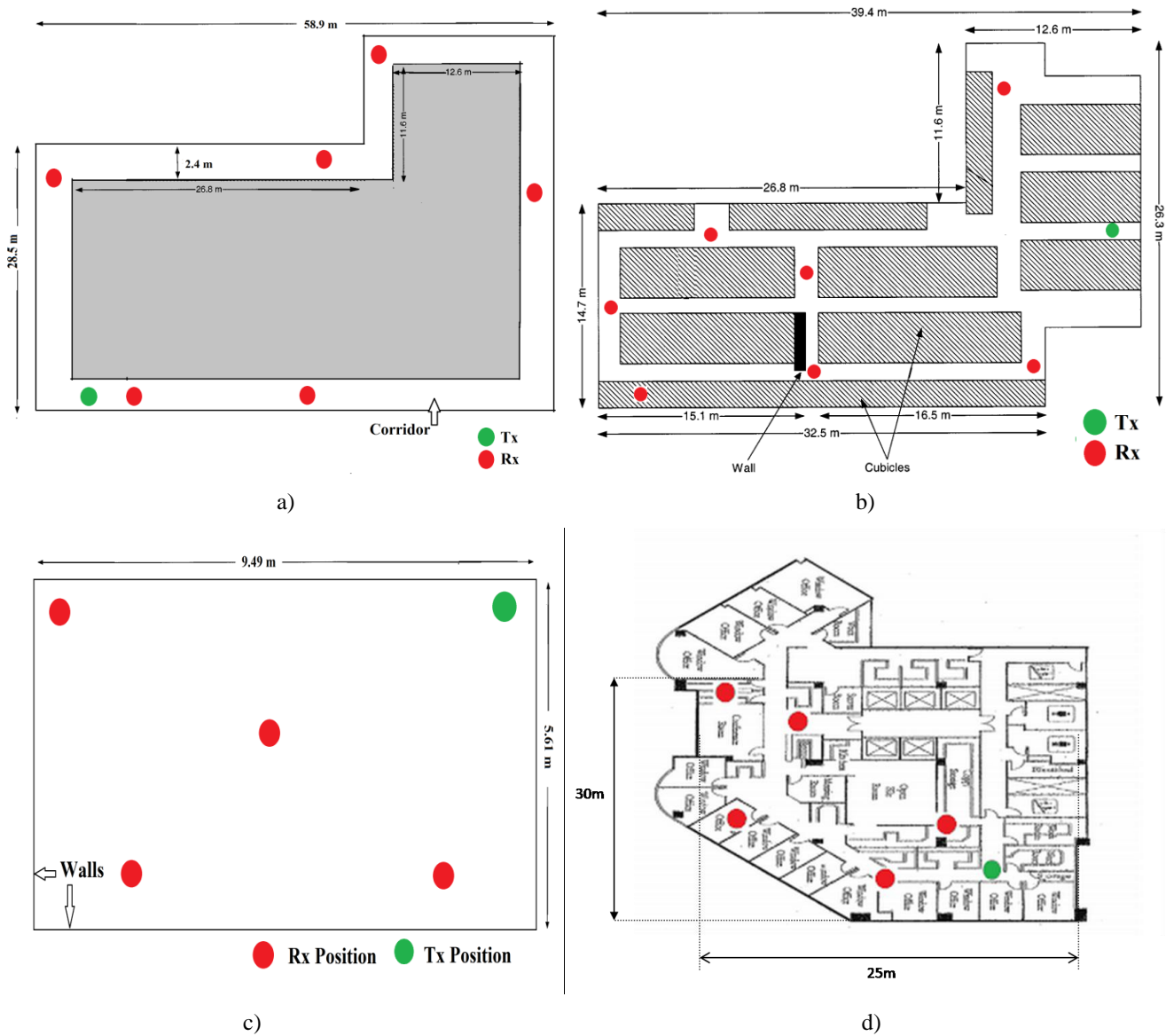

Fig. 7. Layout of the measurment locations. a) 5[th] floor corridor in Engineering IV Building, b) Cubicle Area, c) Faraday Conference Room, d) Silvus Technologies office.